\documentclass[12pt]{article}
\usepackage{a4}
\usepackage{amsmath}
\usepackage{amsfonts}
\usepackage{epsfig}
\def\pa{\partial}

\def\nn{\nonumber}

\def\arctan{{\rm arctan}}
\def\ga{\gamma}
\def\Ga{\Gamma}
\def\si{\sigma}

\def\de{\delta}
\def\ka{\kappa}
\def\al{\alpha}

\def\la{\lambda}

\def\eps{\epsilon}

\def\d{{\rm d}}
\def\cL{{\cal L}}

\def\cT{{\cal T}}

\def\cZ{{\cal Z}}
\def\bC{{\bf C}}
\def\bF{{\bf F}}
\def\bM{{\bf M}}
\def\bN{{\bf N}}
\def\bO{{\bf O}}
\def\bP{{\bf P}}
\def\bu{{\bf u}}
\def\RR{\mathbb R}
\def\ZZ{\mathbb Z}
\def\id{{\,\,{\rm l}\!\!\!1}}

\begin{document}

\begin{flushright}
hep-th/0205166
\end{flushright}

\vskip 2 cm
\begin{center}
{\Large {\bf Brane induced supersymmetry breakdown 
\\[12pt]
and restoration}} 

\vspace{12mm}
{\large
{\bf K.A.\ Meissner ${}^{a,}$\footnote{
\ \ E-mail: Krzysztof.Meissner@fuw.edu.pl}}, 
{\bf H.P.\ Nilles ${}^{b,}$\footnote{
\ \ E-mail: nilles@th.physik.uni-bonn.de}},
{\bf M.\ Olechowski ${}^{a,b,}$\footnote{
\ \ E-mail: olech@th.physik.uni-bonn.de}
}}\\
\vspace{4mm}
{\it ${}^a$ Institute of Theoretical Physics, Warsaw University,} \\
{\it Ho\.za 69, 00--681 Warsaw, Poland.}\\
\vspace{4mm}
{\it ${}^b$ Physikalisches Institut der Universit\"at Bonn,} \\
{\it Nussallee 12, 53115 Bonn, Germany.}\\
\end{center}

\begin{center}
\vspace{10mm}
{\sl Dedicated to Stefan Pokorski on his 60th birthday}
\end{center}

\vspace{10mm}
\begin{abstract}
We investigate the phenomenon of brane induced supersymmetry breakdown
on orbifolds in the presence of a Scherk--Schwarz mechanism. General
consistency conditions are derived for arbitrary dimensions and the
results are illustrated in the specific example of a 5--dimensional
theory compactified on $S^1/\ZZ_2$. This includes a discussion of the
Kaluza--Klein spectrum and the possibility of a brane induced
supersymmetry restoration.
\end{abstract}

\newpage
%%%%%%%%%%%%%%%%%%%%%%%%%%%%%%%%%%%%%%%%%%%%%%%%%%%%%%%%%%%%%%%%%%%%%%
%%%%%%%%%%%%%%%%%%%%%%%%%%%%%%%%%%%%%%%%%%%%%%%%%%%%%%%%%%%%%%%%%%%%%%
%%%%%%%%%%%%%%%%%%%%%%%%%%%%%%%%%%%%%%%%%%%%%%%%%%%%%%%%%%%%%%%%%%%%%%
%%%%%%%%%%%%%%%%%%%%%%%%%%%%%%%%%%%%%%%%%%%%%%%%%%%%%%%%%%%%%%%%%%%%%%
\section{Introduction}

The search for a satisfactory breakdown of supersymmetry is one of the
most important challenges in higher dimensional quantum field theories
and string theories. Mechanisms at our disposal so far are the
Scherk--Schwarz mechanism \cite{Scherk:1978ta,Scherk:1979zr}, orbifold
twists \cite{Rohm:aq,Dixon:jw,Dixon:1986jc} (or more generally spaces
with non trivial holonomy groups \cite{Candelas:en}) as well as brane
induced supersymmetry breaking \cite{Horava:1996vs,Nilles:1997cm}. 
As in all these cases extra dimensions are involved these mechanisms
show some similarities, but there are also decisive differences. One
important difference concerns the question of the possible appearance
of chiral fermions as a consequence of supersymmetry breaking. As the
compactification of higher dimensional supersymmetric theories usually
leads to $N$--extended supersymmetry in $d$=4, we need a mechanism
that breaks supersymmetry to $N=1$ or $N=0$ while allowing for a
chiral fermion spectrum via the mechanism of orbifold twists.

In specific models we are very often confronted with a situation that
a certain combination of the above mentioned mechanisms is at work, a
complicated situation that needs a careful analysis to identify the
most general properties of such a scheme. Also, it has been suggested
that brane induced supersymmetry breakdown is related to the
Scherk--Schwarz mechanism \cite{Antoniadis:1997ic,Antoniadis:1997xk},  
a conjecture based on the fact that supersymmetry broken at a given
brane could be restored by a similar mechanism on a different brane
\cite{Horava:1996vs,Nilles:1998sx,Meissner:1999ja}. Such a 
mechanism has been analysed in detail in \cite{Meissner:1999ja} 
in the framework of the heterotic M--theory of Ho\v rava and
Witten \cite{Horava:1996ma}. As this is a rather complicated set--up
and as some approximations are involved, the explicit calculations
were quite difficult and not that easy to present in a simple way. In
the present paper we would like to discuss the set--up of the combined
action of the three mechanisms in the most general way and illustrate
the schemes in the framework of simple 5--dimensional examples. Our
general formulae are valid also for $d>5$, but explicit solutions are
much harder to obtain. Meanwhile a similar effect has been
investigated in \cite{Bagger:2001qi,Bagger:2001ep} that has some
overlap with the present work. The situation discussed there is,
however, less general than the one considered in the present work 
(and we also differ in some of the explicit formulae).

As these mechanisms are in some sense similar, we start in section 2
with a careful definition of Scherk--Schwarz mechanism and
orbifolding. Section 3 then gives the general conditions for
Scherk--Schwarz mechanism on orbifolds and explains the additional
restrictions as compared to the one on manifolds. In section 4 we
discuss the most general spectrum of the fermion masses on orbifolds
with a Scherk--Schwarz mechanism. Here it is crucial to display the
dependence of these masses on the higher dimensional coordinates to
classify the possibilities for brane induced supersymmetry
breakdown. In section 5 we give solutions for specific simple examples
of interest, and point out some subtleties in the discussion of brane
located mass terms. Section 6 summarizes our main results.

%%%%%%%%%%%%%%%%%%%%%%%%%%%%%%%%%%%%%%%%%%%%%%%%%%%%%%%%%%%%%%%%%%%%%%
%%%%%%%%%%%%%%%%%%%%%%%%%%%%%%%%%%%%%%%%%%%%%%%%%%%%%%%%%%%%%%%%%%%%%%
%%%%%%%%%%%%%%%%%%%%%%%%%%%%%%%%%%%%%%%%%%%%%%%%%%%%%%%%%%%%%%%%%%%%%%
%%%%%%%%%%%%%%%%%%%%%%%%%%%%%%%%%%%%%%%%%%%%%%%%%%%%%%%%%%%%%%%%%%%%%%
\section{Compactification, Scherk--Schwarz mechanism and orbifolding}

To explain the mutual relations among compactification, orbifolding
and the Scherk--Schwarz mechanism it is useful to describe all these
three constructions using the same mathematical language. In this
section we compare definitions of these three phenomena. We use a
one--dimensional example to illustrate all important points.

%%%%%%%%%%%%%%%%%%%%%%%%%%%%%%%%%%%%%%%%%%%%%%%%%%%%%%%%%%%%%%%%%%%%%%
%%%%%%%%%%%%%%%%%%%%%%%%%%%%%%%%%%%%%%%%%%%%%%%%%%%%%%%%%%%%%%%%%%%%%%
\subsection{Compactification}

Let us consider a theory defined in $D$ dimensions. Its action is
given by an integral of an appropriate $D$--dimensional Lagrangian
depending on $D$--dimensional fields 
\begin{equation}
S_D=\int_\bF\d^Dz\cL_D\left(\Phi(z)\right)
\,.
\end{equation}
Such a theory is effectively $d$--dimensional at low energies if
the coordinates $z$ of the $D$--dimensional space $\bF$ can be
split into two sets
\begin{equation}
z^M=\{x^\mu,y^m\}
\,,
\end{equation}
($M=1,\ldots,D$; $\mu=1,\ldots,d$; $m=d+1,\ldots,D$) in such a way
that coordinates $y^m$ describe some $(D-d)$--dimensional compact
space $\bC$. In the simplest case this means that the full space--time
is a product of two factors
\begin{equation}
\bF=\bM\times\bC
\end{equation}
where $\bM$ is non--compact and $d$--dimensional (it should be
just the 4--dimensional Minkowski space in realistic models). 
Integrating over the compact coordinates $y$ one can obtain an
effective $d$--dimensional theory valid for
energies much smaller than the inverse of the length scale
characteristic for the size of $\bC$.

The simplest way to compare the Scherk--Schwarz mechanism and the
ordinary compactification is to consider a case when the compact space
$\bC$ can be obtained from a non--compact covering space $\bN$ using
some group $G$. Let $G$ be a discrete group acting freely on
$\bN$. The action of this group is represented by some operators
$\cT_g$ mapping $\bN$ into itself. For all $g_1,g_2,g_3\in G$ they
satisfy the condition
\begin{equation}
(g_1g_2=g_3)
\quad\Rightarrow\quad
(\cT_{g_1}\cT_{g_2}=\cT_{g_3})
\,.
\label{cTgroup}
\end{equation}
The action of $G$ is free which means that $\cT_g$ has fixed points in
$\bN$ only when $g$ is the identity element of $G$.
We identify points which differ by the action of $\cT_g$ for any 
$g\in G$
\begin{equation}
\cT_g(y) \sim y
\,.
\end{equation}
In other words: we identify two points if they belong to the same orbit
in $\bN$. In this way we obtain a compact space
\begin{equation}
\bN\to\bC=\bN/G
\,.
\end{equation}

But identification of points in the space is not enough. We have also
to demand that ``physics'' at two identified points is the 
same. More precisely: we have to allow only such configurations in the
non compact space $\bN$ for which the contribution to the action from
a given point is the same as from any other point identified with it
(same for each point of a given orbit in the covering space).
This should be true at the quantum level in the full theory but in
order to simplify the notation we write it as the classical level
condition for the Lagrangian at two identified points:
\begin{equation}
\cL\left(\Phi\left(x,\cT_g(y)\right)\right)
=
\cL\left(\Phi(x,y)\right)
\,.
\label{Lperiod}
\end{equation}
Only then the action for the non compact space $\bN$ is equivalent 
to that for the compact space $\bC$ (they differ only by an
unimportant normalization constant).

In the ordinary compactification the above requirement is fulfilled by 
demanding that all the fields have the analogous periodicity property
under $\cT_g$: 
\begin{equation}
\Phi\left(x,\cT_g(y)\right)=\Phi(x,y)
\,.
\end{equation}
This condition is of course sufficient to satisfy eq.\ (\ref{Lperiod})
but in general it is not necessary. It is enough to demand that
fields at $\cT_g(y)$ are related to fields at $y$ by some
transformations: 
\begin{equation}
\Phi\left(x,\cT_g(y)\right)=T_g\Phi(x,y)
\label{PhiPeriod}
\end{equation}
where operations $T_g$ are elements of the 
{\it global}$\,\,$\begin{footnote}
{If $T$ is an element of a local symmetry group than we have a
Hosotani mechanism 
\cite{Hosotani:1983xw,Hosotani:1988bm} 
which is equivalent to gauge symmetry breaking by
nontrivial Wilson lines.}
\end{footnote}$\!$
symmetry group of the theory (which again we write at a classical level
only):
\begin{equation}
\cL\left(T_g\Phi\right)=\cL\left(\Phi\right)
\,.
\label{PhiSS}
\end{equation}

%%%%%%%%%%%%%%%%%%%%%%%%%%%%%%%%%%%%%%%%%%%%%%%%%%%%%%%%%%%%%%%%%%%%%%
%%%%%%%%%%%%%%%%%%%%%%%%%%%%%%%%%%%%%%%%%%%%%%%%%%%%%%%%%%%%%%%%%%%%%%
\subsection{Scherk--Schwarz mechanism}

Now it is easy to define the Scherk--Schwarz mechanism: it is
such compactification for which at least some of twist transformations
$T_g$ are different from identity. The ordinary compactification is
the very special case when $T_g=\id$ for all $g\in G$.

Of course the transformations $T_g$ in the field space can not be
arbitrary. They, similarly to  the transformations $\cT_g$ in the
physical space (\ref{cTgroup}), must respect the group structure of
$G$: 
\begin{equation}
(g_1g_2=g_3)
\quad\Rightarrow\quad
(T_{g_1}T_{g_2}=T_{g_3})
\,.
\label{Tgroup}
\end{equation}
In other words, the transformations $T_g$ must form an appropriate
representation of $G$. 
This is obvious because for every $\{x,y\}$ and $g_3=g_1g_2$ we get
\begin{equation}
T_{g_1}T_{g_2}\Phi(x,y)=T_{g_1}\Phi(x,\cT_{g_2}(y))
=
\Phi(x,\cT_{g_1}\cT_{g_2}(y))=\Phi(x,\cT_{g_3}(y))
=T_{g_3}\Phi(x,y)
\,.
\end{equation}
No additional twists are allowed because $\cT_{g_1}\cT_{g_2}(y)$ and 
$\cT_{g_3}(y)$ denote the same point in the covering space $\bN$ 
(and not two different points which are identified).

The above definitions are quite simple but nevertheless there is some
confusion about the Scherk--Schwarz mechanism in the literature. It
seems that the reason is the following: In both kinds of
compactification the fields $\Phi$ are functions on the non--compact
space $\bN$. In the ordinary compactification they are also functions
on the compact space $\bC$ because of the condition (\ref{PhiPeriod}).
On the other hand, in the presence of some nontrivial Scherk--Schwarz
twists $T_g$, at least some of the fields can not be described by
(single--valued) functions on $\bC$. Instead, they can be described
by sections of some nontrivial fiber bundle with the compact
space $\bC$ as a base space. Of course the structure of that fiber
bundle is not arbitrary, it is determined by the twist operators
$T_g$.

Using the notion of fiber bundles it is possible to define theories on
the compact space $\bC$ even without referring to the non--compact
space $\bN$ (sometimes $\bC$ can not be obtained as $\bN/G$). We use
only $\bC$ and define fields as sections of fiber bundles over
$\bC$. Ordinary compactification corresponds to a case when this fiber
bundle is trivial, i.e. just a product of the field space and the base
space. Using this formalism one can also check when a nontrivial
Scherk--Schwarz mechanism is at all possible. To apply this mechanism
we need a nontrivial fiber bundle. Such bundles exist only when the 
base space ($\bC$ in our case) is a non contractible one.

Let us illustrate our discussion with the simplest possible example,
that of the one dimensional circle $S^1$. 
It can be obtained from the one dimensional real space, $\bN=\RR$, by
using the group of addition of integer numbers, $G=\ZZ$. The $n$--th
element of $\ZZ$ is represented on $\RR$ by the translation by 
$2\pi nR$:
\begin{equation}
\cT_n(y)=y+2\pi nR
\,.
\label{cTn}
\end{equation}
Identifying points which differ by the action of any of these
translations we obtain a fundamental domain of length $2\pi R$ which
can be described by $y\in [y_0,y_0+2\pi R[$ or 
$y\in\,]y_0,y_0+2\pi R]$ 
for arbitrary $y_0$. The interval must be open at one end because
$y=y_0$ and $y=y_0+2\pi R$ describe the same point in the compact
space and should not be counted twice.

The group $\ZZ$ has infinitely many elements but all of them can be
obtained from just one, represented by translation by $2\pi R$. Thus
we need only one independent twist transformation $T$:
\begin{equation}
\Phi(x,y+2\pi R)= T\Phi(x,y)
\,.
\end{equation}
Other transformations are powers of this one: $T_n=T^n$.
Of course $T$ must be a global symmetry of the Lagrangian.

Let us simplify this even further and consider only one real field 
$\phi$. For ordinary compactification $T=\id$, and our theory is
described by real functions on $S^1$: $\phi(y)$ (we drop dependence on
$x$). Non trivial Scherk--Schwarz mechanism is 
obtained e.g. for $T=-1$ (of course 
the Lagrangian must be invariant under $\phi\to-\phi$) and the theory 
can be described in terms of sections of a M\"obius strip.
How can we go back to a description in term of functions?
We need a fundamental domain in the covering space $\RR$.
As we discussed after eq.\ (\ref{cTn}) it can be chosen to be 
$[y_0,y_0+2\pi R[\,$. One can use any $y_0$ but $y_0=-\pi R$ is a
good choice if one wants to use even and odd functions. 
So sections of the
M\"obius strip are represented by single--valued functions on
$[-\pi R,\pi R]$
(we may add the endpoint $y=\pi R$ and define that the value of a
function at $y=\pi R$ is equal to an appropriate limit) 
with additional condition
\begin{equation}
\phi(\pi R)=T\phi(-\pi R) = -\phi(-\pi R)
\,.
\end{equation}
In practical calculations it is usually more convenient to work with
these functions than with sections of the M\"obius strip.
But of course one has to remember that they are single--valued
functions on the interval $I=[-\pi R, \pi R]$, 
{\it but in general are NOT single--valued functions on the circle
$S^1$}.

It is important to remember also that the position of ``the point of
discontinuity'' ($y_0=\pm\pi R$ in the above example), has no real
meaning -- one can not say in a 
meaningful way at which point the M\"obius strip is twisted.
The Scherk--Schwarz mechanism is related to global properties of
the fields and does not distinguish any particular point(s) in the
compact space.

%%%%%%%%%%%%%%%%%%%%%%%%%%%%%%%%%%%%%%%%%%%%%%%%%%%%%%%%%%%%%%%%%%%%%%
%%%%%%%%%%%%%%%%%%%%%%%%%%%%%%%%%%%%%%%%%%%%%%%%%%%%%%%%%%%%%%%%%%%%%%
\subsection{Orbifolding}

Let us now discuss the orbifolding. It is a very important
construction applied in some higher dimensional theories. It can be
used to obtain chiral fermions starting from a model with only non
chiral ones. It has to be contrasted with the compactification which
does not change the chiral structure of the theory to which it is
applied. Nevertheless, using the language introduced in this section,
it is possible to define the orbifolding in a very similar way to that
used to analyse the Scherk--Schwarz compactification. We start with a
space described by a manifold $\bP$ and some discrete group $H$ which
is represented by operations $\cZ_h$ transforming $\bP$ into
itself. We identify points in $\bP$ which differ by the action of
$\cZ_h$ for any $h\in H$ and demand that the fields at such two points
differ by some transformation $Z_h$:
\begin{eqnarray}
\cZ_h(y) 
\!\!\!&\sim&\!\!\!
y
\,,
\\[6pt]
\Phi(x,\cZ_h(y))
\!\!\!&=&\!\!\!
Z_h\Phi(x,y)
\,,
\end{eqnarray}
and all these transformations $Z_h$ must be global symmetries of the
theory. 
The only difference with the compactification is that the group $H$
does not act freely in $\bP$. Some of the transformations $\cZ_h$ have
fixed point in $\bP$ and the resulting space is in general not a
manifold but an orbifold 
\begin{equation}
\bP \to \bO=\bP/H
\,.
\end{equation}
Contrary to the Scherk--Schwarz compactification, there are special
points in the space obtained by orbifolding. The resulting space
(orbifold) is no longer a smooth manifold.

The simplest and very popular example is that of the circle $S^1$
divided by the two element group $\ZZ_2$. The action of the only
nontrivial element of $\ZZ_2$ is represented by the reflection
\begin{equation}
\cZ(y)=-y
\,.
\end{equation}
This operation squares to identity so the same must be true for the
corresponding operation $Z$ in the field space. Thus it is always
possible to choose a basis in which all fields have well defined
parities
\begin{equation}
\Phi\left(x,\cZ(y)\right)=\Phi(x,-y)=Z\Phi(x,y)
\end{equation}
where $Z$ is a diagonal matrix with eigenvalues $\pm1$.

This one--dimensional example is somewhat special. The orbifold
$S^1/\ZZ_2$ is equivalent to a manifold with a boundary (the fixed
points have codimension 1 and can be treated as boundaries).
In general, orbifolds are not equivalent to manifolds with
boundaries.

%%%%%%%%%%%%%%%%%%%%%%%%%%%%%%%%%%%%%%%%%%%%%%%%%%%%%%%%%%%%%%%%%%%%%%
%%%%%%%%%%%%%%%%%%%%%%%%%%%%%%%%%%%%%%%%%%%%%%%%%%%%%%%%%%%%%%%%%%%%%%
%%%%%%%%%%%%%%%%%%%%%%%%%%%%%%%%%%%%%%%%%%%%%%%%%%%%%%%%%%%%%%%%%%%%%%
%%%%%%%%%%%%%%%%%%%%%%%%%%%%%%%%%%%%%%%%%%%%%%%%%%%%%%%%%%%%%%%%%%%%%%
\section{Consistency conditions for Scherk--Schwarz mechanism on
orbifolds}

Let us now discuss the situation when we perform orbifolding and 
Scherk--Schwarz compactification together in one theory. 
Many models of this type, especially 5--dimensional ones, have been
recently proposed in the literature. The orbifolding is necessary to
obtain chiral fermions and also breaks some supersymmetry while the
Scherk--Schwarz mechanism can be used to break the remaining
supersymmetry. We will see that it is quite simple to analyse 
both of these mechanisms simultaneously using the formalism of the
previous section. In the first subsection we present the
consistency conditions which must be fulfilled for a general
Scherk--Schwarz compactification on orbifolds. In the second
subsection we discuss in more detail the important case of the
$S^1/\ZZ_2$ orbifold.

%%%%%%%%%%%%%%%%%%%%%%%%%%%%%%%%%%%%%%%%%%%%%%%%%%%%%%%%%%%%%%%%%%%%%%
%%%%%%%%%%%%%%%%%%%%%%%%%%%%%%%%%%%%%%%%%%%%%%%%%%%%%%%%%%%%%%%%%%%%%%
\subsection{General case}
\label{generalcase}

As we discussed in the previous section, the Scherk--Schwarz
compactification and the orbifolding, despite important 
differences between them, can be described using the same
formalism. Also the situation when both constructions appear
simultaneously is quite straightforward to analyse. We start with a
non compact space $\bN$ and a discrete  
group $F$ acting on it. This group $F$ must be only a little bit more
complicated than in the previous cases. It contains a non trivial
subgroup which acts freely on $\bN$ (and is used to make the resulting
space compact like in the Scherk--Schwarz mechanism) but has also
non trivial elements which have fixed points when acting on $\bN$
(like in orbifolding). Let us denote those different types of elements
of $F$ by $g$ and $h$, respectively. They are represented by
transformations $\cT_g$ and $\cZ_h$. They in fact form one
representation of $F$ and we use different letters only to distinguish
those transformations which have fixed points.

As usually we identify points which differ by the action of any
(combination) of those transformations. We allow also for, in general
non trivial, twists in the field space: 
\begin{eqnarray}
\cT_g(y) &\sim& y
\,,
\label{TZ1}
\\[6pt]
\cZ_h(y) &\sim& y
\,,
\\[6pt]
\Phi(x,\cT_g(y)) &=& T_g\Phi(x,y)
\,,
\\[6pt]
\Phi(x,\cZ_h(y)) &=& Z_h\Phi(x,y)
\,.
\label{TZ4}
\end{eqnarray}
The twist operators $T_g$ and $Z_h$ must of course form a
representation of the group $F$ and must be global symmetries of the
theory. In full analogy to eq.\ 
(\ref{Tgroup}) they have to satisfy the appropriate consistency
conditions also for the ``mixed'' products, e.g.
\begin{equation}
(g_1h_2=h_3)
\quad\Rightarrow\quad
(T_{g_1}Z_{h_2}=Z_{h_3})
\,.
\label{TZgroup}
\end{equation}

Now we can easily compare the Scherk--Schwarz mechanism without and
with orbifolding. In both cases we start with the same non compact
space $\bN$. In the first case we use a group $G$ which free action on
$\bN$ is represented by operators $\cT_g$. The action of that group in
the field space is represented by $T_g$. Then we enlarge the group in
such a way that some of its elements have fixed points when acting on
$\bN$. For definiteness we may chose it to be a direct product: 
$F=G\times H$. The second subgroup $H$ is represented
by some non freely acting operators $\cZ_h$. What is the influence of
the orbifolding group $H$ on the Scherk--Schwarz twists $T_g$? Are
they more restricted or can they be more general? The answer is
obvious: after orbifolding the Scherk--Schwarz twists are more
restricted as compared to the same theory without orbifolding. The
reason is that there are additional consistency conditions of type
shown in eq. (\ref{TZgroup}).

It occurs that those additional consistency conditions can be quite
restrictive. To show this we investigate now the interplay between
the Scherk--Schwarz mechanism and orbifolding in the important case of
the one--dimensional circle.

%%%%%%%%%%%%%%%%%%%%%%%%%%%%%%%%%%%%%%%%%%%%%%%%%%%%%%%%%%%%%%%%%%%%%%
%%%%%%%%%%%%%%%%%%%%%%%%%%%%%%%%%%%%%%%%%%%%%%%%%%%%%%%%%%%%%%%%%%%%%%
\subsection{Scherk--Schwarz mechanism on $\boldsymbol{S^1/\ZZ_2}$
orbifold}

We obtain the orbifold $S^1/\ZZ_2$ by dividing  the real axis $\RR$ by
the group $\ZZ\times\ZZ_2$. It is enough to consider one element of
$\ZZ$ and one element of $\ZZ_2$. The equations (\ref{TZ1}-\ref{TZ4})
take the following form:
\begin{eqnarray}
\cT(y)&=&y+2\pi R
\,,
\label{S1Z2a}\\[6pt]
\cZ(y)&=&-y
\,,
\label{S1Z2b}\\[6pt]
\Phi(x,y+2\pi R)&=&T\Phi(x,y)
\,,
\label{S1Z2c}\\[6pt]
\Phi(x,-y)&=&Z\Phi(x,y)
\,.
\label{S1Z2d}
\end{eqnarray}
Now we want to find the additional consistency conditions of the form
presented in eq.\ (\ref{TZgroup}). There is one such condition and it
follows from the fact that translation $\cT$ and reflection $\cZ$ for
arbitrary $y$ fulfil the condition:
\begin{equation}
\cT\cZ\cT(y)=\cZ(y)
\,,
\end{equation}
from which it follows that
\begin{eqnarray}
TZT\Phi(x,y)
&\!\!=\!\!&
TZ\Phi\left(x,\cT(y)\right)=T\Phi(x,\cZ\cT(y))
\nn\\[6pt]
&\!\!=\!\!&
\Phi(x,\cT\cZ\cT(y))=\Phi(x,\cZ(y))=Z\Phi(x,y)
\,.
\end{eqnarray}
So the operators in the field space must satisfy the relation 
\begin{equation} 
TZT=Z
\,.
\label{TZT}
\end{equation}
We will show below that the above condition puts quite strong
restrictions on the possible form of the twist $T$.

We know that the operator $Z$ must square up to identity so its
eigenvalues must be equal to $1$ or $-1$. Let us start with a basis 
in which the first $n$ eigenvalues of $Z$ are $+1$ and the last $m$
eigenvalues are $-1$. In such a basis $Z$ and $T$ matrices have the form
\begin{equation}
Z=
\left(
\begin{array}{cc}
\id_n & 
\\[6pt]
 & -\id_m
\end{array}
\right)
,\qquad\qquad
T=
\left(
\begin{array}{cc}
A & B
\\[6pt]
C & D
\end{array}
\right)
.
\label{ZTmatrix}
\end{equation}
Multiplying eq.\ (\ref{TZT}) with $T^\dagger$ and using the fact that
$T$ is unitary we get the following condition
\begin{equation}
ZT=T^\dagger Z
\end{equation}
Substituting $Z$ and $T$ in the form (\ref{ZTmatrix}) to this equation
we find that the diagonal blocks, $A$ and $D$, are hermitian while the
off--diagonal ones fulfil the condition $C=-B^\dagger$. Thus we can 
change the basis in two ($n\times n$ and $m\times m$) subspaces in
such a way that $T$ has the form  
\begin{equation}
T=\left(
\begin{array}{cc}
A & B
\\[6pt]
-B^\dagger & D
\end{array}
\right)
\end{equation}
with diagonal and real $A$ and $D$. Now we again
use eq.\ (\ref{TZT}); multiplying it with $Z$ we find
\begin{equation}
(TZ)^2=(TZT)Z=Z^2=\id
\,,\label{TZ2}
\end{equation}
which in terms of the matrices $A$, $B$ and $D$ reads
\begin{eqnarray}
A^2+BB^\dagger&=&\id_n
\,,\label{TZ2a}\\[6pt]
D^2+B^\dagger B&=&\id_m
\,,\label{TZ2b}\\[6pt]
AB-BD&=&0
\,.\label{TZ2c}
\end{eqnarray}
The last equation can be rewritten using the components of the
matrices as
\begin{equation}
B_{ij}\left(A_{ii}-D_{jj}\right)=0
\end{equation}
for all $i=1,\ldots,n$ and $j=1,\ldots,m$. This means that the
elements of the off--diagonal matrix $B$ can be non
zero only in subspaces in which the diagonal matrices ($A$ and $D$)
have equal eigenvalues. So now we can change the basis in such a way
that the matrices $Z$ and $T$ have the following form:
\begin{eqnarray}
Z&=&\left(
\begin{array}{ccccc}
\id_{n_1}&&&&\\
&-\id_{m_1}&&&\\
&&\id_{n_2}&&\\
&&&-\id_{m_2}&\\
&&&&...
\end{array}
\right)
,\\[6pt]
T&=&\left(
\begin{array}{ccccc}
a_1\id_{n_1}&B_1&&&\\
-B_1^\dagger&a_1\id_{m_1}&&&\\
&&a_2\id_{n_2}&B_2&\\
&&-B_2^\dagger&a_2\id_{m_2}&\\
&&&&...
\end{array}
\right)
,
\label{Tblock1}
\end{eqnarray}
where all $a_i$ are different. 
Let us concentrate on the first $(n_1+m_1)\times(n_1+m_1)$ block of
$T$. We can perform two arbitrary changes of basis, one in the $n_1$ 
dimensional subspace and second in the $m_1$ dimensional one, and the
structures of $Z$ and $T$ matrices remain the same. We can use this
freedom to put the $B_1$ matrix in the form (for $n_1\leq m_1$; the
other case can be analysed in an analogous way)
\begin{equation}
B_1=\left(
\begin{array}{cccc}
b_1 & & &\\
& b_2 & & \\
& & \cdot & \\
& & & b_{n_1}\,\,0\,\,0\,...
\end{array}
\right)
,
\end{equation}
with real $b_i$. Now we can use the conditions (\ref{TZ2a}) and
(\ref{TZ2b}). It is easy to see that there are two possibilities:
either $a_1=\pm1$, $B_1=0$ or $n_1=m_1$, $B_1=b_1\id_{n_1}$
with the constant $b_1$ satisfying
\begin{equation}
a_1^2+b_1^2=1
\,.
\end{equation}
We can perform now the last change of the basis: we permute
appropriately the coordinates in subspaces with $n_i>1$. Now the
matrices $Z$ and $T$ have their final form:
\begin{eqnarray}
Z&=&\left(
\begin{array}{cccc}
\sigma_3 & & &\\
& \sigma_3 & & \\
& & \cdot & \\
& & & \pm\id_{|n-m|}
\end{array}
\right)
,\\[6pt]
T&=&\left(
\begin{array}{cccc}
R(2\pi\alpha_1) & & &\\
& R(2\pi\alpha_2) & & \\
& & \cdot & \\
& & & I_{|n-m|}
\end{array}
\right)
,
\label{Tfinalform}
\end{eqnarray}
where $I_{|n-m|}$ is a diagonal matrix of dimension $|n-m|$ 
with diagonal entries equal $\pm1$ while
$R(2\pi\alpha_i)$ is a matrix describing rotation by
an angle $2\pi\alpha_i$: 
\begin{equation}
R(2\pi\alpha_i)=
\left(
\begin{array}{rr}
\cos(2\pi\alpha_i) & -\sin(2\pi\alpha_i)
\\[6pt]
\sin(2\pi\alpha_i) & \cos(2\pi\alpha_i) 
\end{array}
\right)
.
\end{equation}
Observe that now $T$ is block--diagonal with only 2-- and
1--dimensional subspaces. If any of the dimensions $n_i$ in the form
(\ref{Tblock1}) is bigger than 1 then the corresponding
$2n_i\times2n_i$ subspace decomposes to $n_i$ 
block--diagonal entries with the same rotation angle $2\pi\alpha$.

We see that the possible Scherk--Schwarz mechanism in the
case of the one dimensional orbifold $S^1/\ZZ_2$ is quite restricted. 
The only allowed twists are: the rotations in two--dimensional
subspaces consisting of one field which is even under the $\ZZ_2$
parity and one field which is odd; and the change of sign of some
fields which are not rotated. It should be stressed that
(contrary to some claims in the literature \cite{Bagger:2001qi})
orbifolding of the circle does not open any possibilities for
generalizing the Scherk--Schwarz mechanism. The situation is just
opposite: additional, quite strong constraints must be fulfilled. One
can no longer use any arbitrary global symmetry of the theory for the
twists, only  twists of the form (\ref{Tfinalform}) can be
consistently used. 
In particular one can not ``generalize'' the Scherk--Schwarz mechanism
by allowing for extra discontinuities of the fields at the fixed
points of the orbifold. Any discontinuities, as well as other local
features of the fields, are determined by appropriate equations of
motion. As we have already stressed, the Scherk--Schwarz mechanism
determines only the global properties of the fields.

%%%%%%%%%%%%%%%%%%%%%%%%%%%%%%%%%%%%%%%%%%%%%%%%%%%%%%%%%%%%%%%%%%%%%%
%%%%%%%%%%%%%%%%%%%%%%%%%%%%%%%%%%%%%%%%%%%%%%%%%%%%%%%%%%%%%%%%%%%%%%
%%%%%%%%%%%%%%%%%%%%%%%%%%%%%%%%%%%%%%%%%%%%%%%%%%%%%%%%%%%%%%%%%%%%%%
%%%%%%%%%%%%%%%%%%%%%%%%%%%%%%%%%%%%%%%%%%%%%%%%%%%%%%%%%%%%%%%%%%%%%%
\section{Fermion spectrum on $\boldsymbol{S^1/\ZZ_2}$ with
Scherk--Schwarz mechanism and mass terms}

Originally the Scherk--Schwarz mechanism
\cite{Scherk:1978ta,Scherk:1979zr} was used to break
supersymmetry. The masses of all levels of the Kaluza--Klein tower 
(especially for gravitini) were
shifted by a constant. It is interesting to check how mass levels are 
changed by the Scherk--Schwarz mechanism on orbifolds. We will
concentrate on the compactification from 5 to 4 dimensions on
$S^1/\ZZ_2$.

%%%%%%%%%%%%%%%%%%%%%%%%%%%%%%%%%%%%%%%%%%%%%%%%%%%%%%%%%%%%%%%%%%%%%%
%%%%%%%%%%%%%%%%%%%%%%%%%%%%%%%%%%%%%%%%%%%%%%%%%%%%%%%%%%%%%%%%%%%%%%
\subsection{Kinetic Lagrangian in 5 dimensions}

Many 5--dimensional models using the compactification on
$S^1/\ZZ_2$  have been discussed recently in the literature.
But instead of choosing any specific model of this type we will
consider a rather general situation. Thus our analysis can be used
to investigate many different, not only 5--dimensional theories, by
specifying some parameters.

Let us consider one 5--dimensional fermion field.
Usually it is described by a pair of spinors satisfying the symplectic
Majorana condition
\begin{equation}
(\la^i)^*=C_5\eps_{ij}\la^j
\end{equation}
where $C_5$ is the 5--dimensional charge conjugation matrix. In the
case of the $S^1/\ZZ_2$ compactification those spinors have the
following $\ZZ_2$ parity properties
\begin{eqnarray}
\la^1(-y)
\!\!\!&=&\!\!\!
+\Ga^5\la^1(y)
\,,\\[6pt]
\la^2(-y)
\!\!\!&=&\!\!\!
-\Ga^5\la^2(y)
\,.
\end{eqnarray}
But it is not very convenient to use these 5--dimensional symplectic
Majorana spinors. We are interested in the compactified, effectively
4--dimensional theory. So let us define two new spinors via the
relations
\begin{eqnarray}
\psi^1
\!\!\!&=&\!\!\!
\frac{1+\Ga^5}{2}\la^1 - \frac{1-\Ga^5}{2}\la^2
\,,\\[6pt]
\psi^2
\!\!\!&=&\!\!\!
\frac{1-\Ga^5}{2}\la^1 + \frac{1+\Ga^5}{2}\la^2
\,.
\end{eqnarray}
It is easy to check that these new spinors fulfil the 4--dimensional
Majorana condition
\begin{equation}
(\psi^i)^*=C_4\psi^i
\end{equation}
(with $C_4$ being the 4--dimensional charge conjugation matrix) and 
have well defined parities under $\ZZ_2$:
\begin{eqnarray}
\psi^1(-y)
\!\!\!&=&\!\!\!
+\psi^1(y)
\,,\label{psi1parity}
\\[6pt]
\psi^2(-y)
\!\!\!&=&\!\!\!
-\psi^2(y)
\,.\label{psi2parity}
\end{eqnarray}
Then the 5--dimensional kinetic term for our spinor $\la^i$ can be
rewritten in terms of $\psi^i$
\begin{equation}
-\frac12\overline{\la^i}\Ga^M\pa_M\la_i
=
-\frac12\left[
\overline{\psi^1}\ga^\mu\pa_\mu\psi^1
+\overline{\psi^2}\ga^\mu\pa_\mu\psi^2
-\overline{\psi^1}\pa_y\psi^2
+\overline{\psi^2}\pa_y\psi^1
\right]
\end{equation}
where we dropped eventual couplings to gauge bosons. We add also 
direct mass terms for the fermions. To be as general as possible we
allow for the $y$ dependence in these mass terms. There can be two
kinds of such mass terms: even and odd under the $\ZZ_2$ parity:
\begin{equation}
m_\pm(-y)=\pm m_\pm(y)
\,.
\end{equation}
Taking into account the parity properties of $\psi^i$ 
(\ref{psi1parity}), (\ref{psi2parity}) we get the following
$\ZZ_2$ invariant kinetic Lagrangian
\begin{eqnarray}
\cL_{\rm kin}
=
-\frac12\Big[
&\!\!\!+\!\!\!&
\overline{\psi^1}\ga^\mu\pa_\mu\psi^1
+\overline{\psi^2}\ga^\mu\pa_\mu\psi^2
\nn\\[6pt]
&\!\!\!-\!\!\!&
\overline{\psi^1}\pa_y\psi^2
+\overline{\psi^2}\pa_y\psi^1
-m_+\left(\overline{\psi^1}\psi^1+\overline\psi^2\psi^2\right)
-m_-\left(\overline{\psi^1}\psi^2+\overline\psi^2\psi^1\right)
\Big]
.\ \ \ 
\label{Lkin}
\end{eqnarray}
The last four terms in the square bracket will give effective
4--dimensional mass terms after compactification (integration over the
5--th coordinate $y$).

One could think about further generalization of the above Lagrangian
by allowing for two independent $\ZZ_2$--even mass terms, one for
$\psi^1$ and another for $\psi^2$. This could be an option for
two independent spinors $\psi^i$ but not in models discussed here. The
spinors $\psi^1$ and $\psi^2$ are related. They are just
different components of one 5--dimensional spinor. Before orbifolding,
all the interactions for $\psi^2$ are strictly determined by those for
$\psi^1$ simply by 5--dimensional Lorentz invariance. After
orbifolding there is only one quantum number which differentiate
between $\psi^1$ and $\psi^2$; the $\ZZ_2$ parity which is even for
one field and odd for the other. But this does not influence terms
quadratic in any of these fields because such terms are $\ZZ_2$--even
anyway. So, in 5--dimensional theories compactified on $S^1/\ZZ_2$ the
$\ZZ_2$--even mass term $m_+$ should be the same for both 
fermions.\footnote{
The authors of refs.\ \cite{Bagger:2001qi,Bagger:2001ep} also consider 
Lagrangians which do not agree with this conclusion. In particular in
eq.\ (3.1) in \cite{Bagger:2001ep} they assume that there is a delta
like mass term for $\psi^1$ but not for $\psi^2$. It is unclear how
such a situation could be realized in a 5--dimensional model.
}
Similar conclusions can be obtained also for higher dimensional
theories. One common $m_+$ mass term appears e.g. in the case of the
11--dimensional heterotic M--theory \cite{Meissner:1999ja} which is a
practical realization of the situation discussed in this paper.

The rest of this section is devoted to the analysis of the effective
4--dimensional spectrum of fermions coming from this 5--dimensional
Lagrangian (\ref{Lkin}) after compactification on $S^1/\ZZ_2$ with
possible Scherk--Schwarz twists.

A few remarks about the possible origin of such a Lagrangian and 
$y$--dependent mass terms are in order. The even mass terms, constant
or located at the fixed points of the orbifold, can be explicitly
present in the model under consideration. Other mass terms can not 
appear directly because they are not allowed by the symmetries of the
theory (e.g.\ the direct odd mass term is forbidden by $\ZZ_2$
parity). But they can appear indirectly when some fields develop non
zero vacuum expectation values 
(VEVs) which break those symmetries. More generally, the Lagrangian
(\ref{Lkin}) should be treated as a part of an effective Lagrangian
obtained in a given theory after some operations. Such operations can
be e.g.: taking into account non zero VEVs of the background fields,
redefinitions of fields, reduction from higher dimensions (if we start
with a theory which is more than 5--dimensional),
changing from a possible warped metric to an effective flat one etc.
Our analysis can be applied to all situations when after all necessary
redefinitions we can get the Lagrangian in the form of (\ref{Lkin}). 
That Lagrangian is written for a spin 1/2 fermion but it can be also
easily generalized to the case of spin 3/2 (we have to add two
gamma matrices between $\overline{\psi^i}$ and $\psi^j$ in an appropriate
way). So our results are valid also for the very interesting case of
gravitini in supersymmetric models.

A very good example of the above mentioned redefinitions is that of the
heterotic M--theory. We analysed the massless gravitino in such a model
in the presence of brane located gaugino condensates in our previous
paper \cite{Meissner:1999ja}. In this model it is necessary to perform
several field redefinitions to take into account six extra dimensions
compactified on a Calabi--Yau manifold. Some redefinitions are
connected to the fact that the background matric is warped. In the
effective 5--dimensional Lagrangian we obtained two types of masses
for the gravitino field, analogous to those present in (\ref{Lkin}).
Both are generated by non zero VEVs of some components of the 4-th
rank tensor field $G_{ABCD}$ present in the 11--dimensional
supergravity. VEV of $G_{11abc}$ gives an even mass term in
5--dimensions while VEV of $G_{a\bar ab\bar b}$ gives an odd one
($a,b$ and $\bar a,\bar b$ are, 
respectively, holomorphic and anti--holomorphic coordinates on the
Calabi--Yau manifold). Thus we see that the $\ZZ_2$ even and odd,
coordinate--dependent mass terms can quite naturally appear in
higher dimensional models.

%%%%%%%%%%%%%%%%%%%%%%%%%%%%%%%%%%%%%%%%%%%%%%%%%%%%%%%%%%%%%%%%%%%%%%
%%%%%%%%%%%%%%%%%%%%%%%%%%%%%%%%%%%%%%%%%%%%%%%%%%%%%%%%%%%%%%%%%%%%%%
\subsection{Mass eigenstate equations and Scherk--Schwarz boundary
conditions}

Before we look for the spectrum of fermions which can be obtained from
the Lagrangian (\ref{Lkin}) we have to specify the properties of the
fields under the Scherk--Schwarz twist. In the previous section we
have proved that the most general twist can be decomposed into
rotations in 2--dimensional subspaces, each consisting of one even and
one odd field. The two Majorana fermions $\psi^1$ and $\psi^2$ form
such a 2--dimensional subspace. In principle it is possible that
$\psi^1$ and $\psi^2$ belong to two different such subspaces if there
are more fields with appropriate quantum numbers (remember that any
Scherk--Schwarz twist must be a global symmetry of the theory so it
can not mix arbitrary fields). In such a case one should consider
$\psi^1$, $\psi^2$ to be vectors and $m_+$, $m_-$ to be matrices in
some type of a flavor space. However we are not going to consider here
such a complication especially because it is not important for the
most interesting case of the gravitino in supersymmetric models. We
concentrate on a 2--dimensional subspace for which the twists are
given by\footnote{
$T$ of this form may be an element of the $SU(2)_R$
automorphism group of the $d=5$ supersymmetry.
}
\begin{equation}
Z=
\left(
\begin{array}{cr}
1&0
\\[6pt]
0&-1
\end{array}
\right)
,\qquad
T=
\left(
\begin{array}{cr}
\cos(2\pi\al)&-\sin(2\pi\al)
\\[6pt]
\sin(2\pi\al)&\cos(2\pi\al)
\end{array}
\right)
,
\end{equation}
in the basis $\Phi=(\psi^1,\psi^2)^T$. 
In this case the twist condition (\ref{S1Z2c}) reads
\begin{equation}
\left(
\begin{array}{c}
\psi^1(x,\pi R)
\\[6pt]
\psi^2(x,\pi R)
\end{array}
\right)
=
\left(
\begin{array}{c}
\cos(2\pi\al)\psi^1(x,-\pi R)-\sin(2\pi\al)\psi^2(x,-\pi R)
\\[6pt]
\sin(2\pi\al)\psi^1(x,-\pi R)+\cos(2\pi\al)\psi^2(x,-\pi R)
\end{array}
\right)
.
\label{psitwist}
\end{equation}
When the twist parameter $\al$ is equal to zero, we have the standard 
compactification in which $\psi^1(y)$ and $\psi^2(y)$ are
(periodic) functions on the circle.

Now we are ready to analyse the spectrum. First we decompose the 
5--dimensional fields $\psi^i$ in the following way:
\begin{equation}
\left(
\begin{array}{c}
\psi^1(x,y)
\\[6pt]
\psi^2(x,y)
\end{array}
\right)
=
\sum_n\chi_n(x)
\left(
\begin{array}{c}
u_n^1(y)
\\[6pt]
u_n^2(y)
\end{array}
\right)
\,.
\label{modes}
\end{equation}
We are looking for such a decomposition for which $\chi_n(x)$ is the
$n$--th 4--dimensional Majorana fermion with a definite masses $M_n$.
The vector of functions $(u_n^1(y),u_n^2(y))^T$ describes the shape 
of this $n$--th mass eigenstate in the 5--th dimension.   
We need both components because in
general mass eigenstates do not have definite parities. 
Substituting this decomposition
into the Lagrangian (\ref{Lkin}) we find that the functions $u_n^i(y)$
must satisfy the following differential equations
\begin{eqnarray}
\frac{\pa u_n^1(y)}{\pa y}+[M_n-m_+(y)]u_n^2(y)-m_-(y)u_n^1(y)
&\!\!\!=\!\!\!&0
\,,\label{diff1}
\\[6pt]
\frac{\pa u_n^2(y)}{\pa y}-[M_n-m_+(y)]u_n^1(y)+m_-(y)u_n^2(y)
&\!\!\!=\!\!\!&0
\,.\label{diff2}
\end{eqnarray}
They can be rewritten in a more compact form as
\begin{equation}
\frac{\pa \bu_n(y)}{\pa y}
+[M_n-m_+(y)]i\si_2\bu_n(y)-m_-(y)\si_3\bu_n(y)=0
\label{diffmatrix}
\end{equation}
where $\si_i$ are Pauli matrices in a space in which the even and odd
components form a vector $\bu_n$. 
The equations alone are not enough, we have to specify also the
boundary conditions. Parity properties of the fields determine the
boundary condition at $y=0$:
\begin{equation}
\left(\begin{array}{c}
u_n^1(0)
\\[6pt]
u_n^2(0)
\end{array}\right)
=
\left(\begin{array}{c}
c_n
\\[6pt]
0
\end{array}\right)
\label{boundcond1}
\end{equation}
where $c_n$ are constants which should be adjusted in order to have the
correct normalization of the 4--dimensional fields. The boundary
condition at $y=\pm\pi R$ can be obtained from eq.\ (\ref{psitwist}).
Substituting expansion (\ref{modes}) into (\ref{psitwist}) and using
the parity properties of $u_n^i(y)$ we get:
\begin{equation}
\frac{u_n^2(\pi R)}{u_n^1(\pi R)}
=
\tan(\pi\al)
\,.
\label{boundcond2}
\end{equation}
The masses and shapes of the 4--dimensional modes can in principle be
found by solving the above differential equations (\ref{diff1}), 
(\ref{diff2}) with the boundary conditions given by
(\ref{boundcond1}), (\ref{boundcond2}). It is possible to simplify this
problem if we are interested only in the masses. To this end we
consider only the ratio of the odd component to the even component of
the wave function: $t_n(y)=u_n^2(y)/u_n^1(y)$. The differential
equation for this function decouples from that for the other
independent combination of $u_n^2(y)$ and $u_n^1(y)$ and reads
\begin{equation}
\frac{\pa t_n(y)}{\pa y}
=
[M_n-m_+(y)]\left(1+t_n^2(y)\right)-2m_-(y)t_n(y)
\,.
\label{difft}
\end{equation}
The appropriate boundary conditions
\begin{eqnarray}
t_n(0)&\!\!\!=\!\!\!&0
\,,\label{boundcond1t}
\\[6pt]
t_n(\pi R)&\!\!\!=\!\!\!&\tan(\pi\al)
\,,\label{boundcond2t}
\end{eqnarray}
can be used to obtain the discrete spectrum of masses $M_n$.
Unfortunately for arbitrary mass terms
$m_\pm(y)$ it is not possible to find the solutions either for
$t_n(y)$ or for the separate components $u_n^i(y)$ in a closed form.

%%%%%%%%%%%%%%%%%%%%%%%%%%%%%%%%%%%%%%%%%%%%%%%%%%%%%%%%%%%%%%%%%%%%%%
%%%%%%%%%%%%%%%%%%%%%%%%%%%%%%%%%%%%%%%%%%%%%%%%%%%%%%%%%%%%%%%%%%%%%%
%%%%%%%%%%%%%%%%%%%%%%%%%%%%%%%%%%%%%%%%%%%%%%%%%%%%%%%%%%%%%%%%%%%%%%
%%%%%%%%%%%%%%%%%%%%%%%%%%%%%%%%%%%%%%%%%%%%%%%%%%%%%%%%%%%%%%%%%%%%%%
\section{Fermion spectrum for some types of models}

In this section we discuss some situations when exact solutions
can be found or when at least some important features of the solutions
can be analysed.

%%%%%%%%%%%%%%%%%%%%%%%%%%%%%%%%%%%%%%%%%%%%%%%%%%%%%%%%%%%%%%%%%%%%%%
%%%%%%%%%%%%%%%%%%%%%%%%%%%%%%%%%%%%%%%%%%%%%%%%%%%%%%%%%%%%%%%%%%%%%%
\subsection{Arbitrary $\boldsymbol{m_+(y)}$ with vanishing
$\boldsymbol{m_-}$} 
\label{mplus}

The situation is very simple when the odd mass term is absent:
$m_-(y)=0$. Then the equations for the modes can be easily
solved. Using the form (\ref{diffmatrix}) we immediately find
\begin{equation}
\bu_n(y)
=
\exp\left\{-i\si_2\int_0^y\d s\,[M_n-m_+(s)]\right\}
\bu_n(0)
\,.
\label{uvector}
\end{equation}
Observe that the above exponent is just equal to the rotation matrix
with the rotation angle given by the integral of $[M_n-m_+]$. 
Thus the mass eigenstates are given by
\begin{equation}
\left(
\begin{array}{c}
u_n^1(y)
\\[6pt] 
u_n^2(y)
\end{array}
\right)
=
\frac{1}{\sqrt{\pi R}}
\left(
\begin{array}{c}
\cos\left[M_ny-\int_0^y\d s\,m_+(s)\right]
\\[6pt]
\sin\left[M_ny-\int_0^y\d s\,m_+(s)\right]
\end{array}
\right)
,
\label{un}
\end{equation}
where we used the boundary condition (\ref{boundcond1}) at $y=0$.
It is very easy to solve also the boundary condition
(\ref{boundcond2}) at $y=\pi R$; the masses $M_n$ must satisfy
the following equality:
\begin{equation}
\tan\left(\int_0^{\pi R}\d y\,[M_n-m_+(y)]\right)
=\tan(\pi\al)
\,,
\end{equation}
hence, they are given by a simple formula
\begin{equation}
M_n=\frac{n+\al+\al_+}{R}
\label{Mn}
\end{equation}
where $\al$ is the Scherk--Schwarz twist parameter and $\al_+$ is
defined by
\begin{equation}
\al_+=\frac{1}{\pi}\int_0^{\pi R}\d y\,m_+(y)
\,.
\label{alphaplus}
\end{equation}
From the above formulae we can see that the Scherk--Schwarz twist
parameter $\al$ and the integrated 5--dimensional, $\ZZ_2$--even mass
term $\al_+$ 
have exactly the same influence on the 4--dimensional mass
eigenvalues. They both shift the masses of all the standard
Kaluza--Klein levels by a constant: $\al/R$ and $\al_+/R$,
respectively.

From eq.\ (\ref{Mn}) it is obvious that there are several
possibilities when those two effects produce no net effect leaving the
masses of the KK states unchanged. In the case of a 
gravitino field in a supersymmetric model this corresponds to unbroken
supersymmetry. This happens when the mass term and the
Scherk--Schwarz parameter satisfy the condition
\begin{equation}
\al+\frac{1}{\pi}\int_0^{\pi R}\d y\,m_+(y) 
=
k \in \ZZ
\qquad {\rm\ \ for\ \ }m_-(y)=0
\,.
\label{unchangedKK}
\end{equation}
This of course does not mean that the Scherk--Schwarz mechanism
is equivalent to arbitrary mass terms satisfying the above equations. 
As we have already stressed, the Scherk--Schwarz mechanism is related
to global properties of the fields and not to their local behavior. 
The Scherk--Schwarz twist parameter appears directly only in the mass
formula. On the other hand the mass term $m_+(y)$ enters explicitly
also the equations determining the shapes of the modes. And those
shapes can be important for example if one considers interactions with
other fields. In fact the Scherk--Schwarz mechanism is equivalent to
the mechanism of adding a mass term only if this mass term is
$\ZZ_2$--even and constant, and for sure 
not when the mass terms are localized at the fixed points of the
orbifold (such delta--like localization of the mass terms is quite
typical for models considered in the literature).

Let us now discuss the possibility of vanishing fermion (gravitino)
mass. The simplest possibility occurs when $k=0$ in eq.\
(\ref{unchangedKK}). In such a case the effects due to the
Scherk--Schwarz twist and the even mass term cancel exactly against
each other in the mass formula (\ref{Mn}) for each KK 
level (this was observed in ref.\ \cite{Bagger:2001qi} for 
delta--like mass terms). A different
interesting situation occurs when the sum of $\al$ and $\al_+$ is a
non zero integer. The structure of the whole KK tower remains
unchanged but the masses of all individual states do change. If we
consider a smooth increase of parameters $\al$ and $\al_+$ from zero
to their final values: the initially massless mode gets non zero mass
while one of the massive modes becomes massless.

This crossing of levels can not occur if $\al,\al_+\ll1$. At least one
of these parameters must be comparable to 1. The Scherk--Schwarz twist
parameter $\al$ can have values only in the range $[-1/2,1/2]$. The
reason is that, as we shown in the previous section, the
Scherk--Schwarz twist in the case of $S^1/\ZZ_2$ is just a rotation by
and angle $2\pi\al$ and the boundary condition (\ref{boundcond2}) 
depends only on $\tan(\pi\al)$. So, $\al$ and $(\al+n)$ describe in
fact exactly the same model. The situation with the mass term $m_+(y)$
and its (normalized) integral $\al_+$ is different. Two models in
which the value of $\al_+$ differs by an integer are really
different. But one should be careful. The crossing of levels can
occur when the average of $m_+(y)$ over the 5--th coordinate satisfies
\begin{equation}
\frac{
\int_0^{\pi R}\d y\,m_+(y)}{\int_0^{\pi R}\d y}
={\cal O}\left(\frac{1}{R}\right)
\end{equation}
and it is necessary to check whether this is still in the range of
validity of used approximations and/or assumptions.

%%%%%%%%%%%%%%%%%%%%%%%%%%%%%%%%%%%%%%%%%%%%%%%%%%%%%%%%%%%%%%%%%%%%%%
%%%%%%%%%%%%%%%%%%%%%%%%%%%%%%%%%%%%%%%%%%%%%%%%%%%%%%%%%%%%%%%%%%%%%%
\subsection{Comments on delta like terms on orbifolds}

In many models discussed in the literature the even mass terms
$m_+(y)$ have the form of Dirac delta sources located at the branes
(fixed points of orbifolds used in those models). Such terms are
sometimes taken improperly into account, so let us make some
comments to clarify this issue. There is a problem which quite often
appears in the literature, namely  when one has to multiply $\de(y)$
by functions vanishing at $y=0$, e.g.\ functions odd in $y$. 
The naive, and incorrect, way is to
assume that all such products are zero because $\de(y)\ne0$ only at
the point for which the other functions do vanish. To clarify this 
we start with reminding the obvious fact that the Dirac delta
``function'' is not a function but a distribution. So one 
should treat it consistently as a distribution or as a limit of some
appropriate functions. Using any of these approaches one can easily
solve the above mentioned problem of multiplying $\de(y)$ by odd
functions. For arbitrary set of (not necessary different) functions
$g_i(y)$ odd in $y$ the following relations hold:
\begin{eqnarray}
\de(y)\prod_{i=1}^{2n+1}g_i(y)
&\!\!\!=\!\!\!&
0
\,,
\\[6pt]
\de(y)\prod_{i=1}^{2n}g_i(y)
&\!\!\!=\!\!\!&
\frac{1}{2n+1}\de(y)\lim_{y\to0}\prod_{i=1}^{2n}g_i(y)
\,.
\end{eqnarray}
From the last equation it follows in particular that the delta like
mass sources couple not only to even but also to odd parity fields
because the odd fields can have jumps at $y=0$ so also nonzero limits
for $y\to0$. And there is no obvious way to forbid such couplings by
some additional symmetries because the odd and even fields are just
components of one 5--dimensional field with definite quantum
numbers.
Because of that we have a common, $\ZZ_2$--even mass term $m_+$ for
both fields, $\psi^1$ and $\psi^2$, in the kinetic Lagrangian
(\ref{Lkin}). Such situation is realized e.g.\ for the gravitino mass
terms in the heterotic M--theory \cite{Meissner:1999ja}.

Another kind of problems can appear when a delta like mass term has
large magnitude. From eq.\ (\ref{un}) we see that the eigenfunctions
in the case of vanishing $m_-$ correspond to a unit vector rotating in
the ($u^1$, $u^2$) space with the $y$--dependent phase angle given by 
\begin{equation}
\varphi(y)=\left[M_ny-\int_0^y\d s\,m_+(s)\right]
\,.
\label{phase}
\end{equation}
Let us discuss the behavior of this angle close to $y=0$. We
regularize $\de(y)$ by some functions $f_\eps(y)$ which integrate to 1
and have support for $|y|<\eps$ with $\eps\to0$. The mass
term is approximated by $m_+(y)=cf_\eps(y)$. For very small
$y$ we can neglect the $M_ny$ contribution in eq.\ (\ref{phase}). 
Then for each function $f_\eps$ the phase at $y=\eps$ is given by:
\begin{equation}
\varphi_\eps(\eps)=-\int_0^\eps\d y\,cf_\eps(y)=-\frac{c}{2}
\,.
\label{phaseeps}
\end{equation}
The phase is the same for all $\eps$ so it has this value also in the
limit $\eps\to0$. Therefore infinitesimally close to the brane the
state vector satisfies the condition
\begin{eqnarray}
u^1_n(\eps)
\!\!\!&=&\!\!\!
\cos\left(\frac{c}{2}\right)u_n^1(0)
,\label{uu1}
\\[6pt]
u^2_n(\eps)
\!\!\!&=&\!\!\!
-\sin\left(\frac{c}{2}\right)u_n^1(0)
.
\label{uu2}
\end{eqnarray}
The values of $c$ which differ by multiples of $4\pi$ give the same
values of $u^1(\eps)$ and $u^2(\eps)$. They give however different
behavior of the state vector over the interval  $y\in[0,\eps]$ for
$\eps\ne0$. The state vector corresponding to a bigger value of $c$
rotates more times by the full angle $2\pi$ between points $y=0$ and
$y=\eps$. Thus, the equations (\ref{uu1}), (\ref{uu2}) taken at
$\eps=0$ are alone not enough to describe a given eigenstate. For
small values of the magnitude $c$ the eigenstate may be described by
functions $u^1(y)$ and $u^2(y)$ which are just discontinuous at
$y=0$. But for large values of $c$ a more careful treatment is
necessary.

The last remark on delta--like terms on orbifolds concerns their
normalization. Performing the integral in eq. (\ref{phaseeps}) we get
only $-c/2$ and not $-c$ because we integrate only over the ``half''
of the delta's support, that for positive $y$. This is the so called 
down--stairs approach to 
the $S^1/\ZZ_2$ orbifold in which we integrate over the interval
$y\in[0,\pi R]$ using the additional prescription that Dirac deltas
located at the fixed points give after integration $1/2$ instead of
1. In the up--stairs approach one integrates over full circle $S^1$
but the final result must be divided by 2.

%%%%%%%%%%%%%%%%%%%%%%%%%%%%%%%%%%%%%%%%%%%%%%%%%%%%%%%%%%%%%%%%%%%%%%
%%%%%%%%%%%%%%%%%%%%%%%%%%%%%%%%%%%%%%%%%%%%%%%%%%%%%%%%%%%%%%%%%%%%%%
\subsection{Constant, simultaneously non zero, $\boldsymbol{m_+}$ and
$\boldsymbol{m_-}$}

Let us now turn to more complicated cases of the Scherk--Schwarz
mechanism on orbifolds, when the odd mass term in the Lagrangian
(\ref{Lkin}) is different from zero. Now it is not possible to solve
the differential equation for the mass eigenstates (\ref{diffmatrix})
by simple integration and exponentiation (as we did to get eq.\
(\ref{uvector}) in the case of $m_-=0$). The reason is that matrices
$\si_2$ and $\si_3$ do not commute and they are multiplied by
functions of $y$. The formal solution of 
(\ref{diffmatrix}) involves an appropriate ordering operator and not
just the ordinary exponential function. Because of that, it is not
possible to write the mass eigenstates, given by solutions of
(\ref{diffmatrix}), in an explicit, closed form. In the case of
arbitrary $m_-(y)$ it is also not possible to solve explicitly eq.\
(\ref{difft}) which determines the masses eigenvalues (without
determining the shapes of the states). 
But there are some simple cases in which we can solve some
of the equations or at least be able do find some properties of the
solutions on which we concentrate in the rest of this section.

One of the cases when it is possible to find explicitly the
eigenstates is when both mass terms $m_+$ and $m_-$ are 
constant. Then the solution to (\ref{diffmatrix}) is given by
\begin{equation}
\bu_n(y)
=
\exp\left\{\left(
-i\si_2[M_n-m_+]+\si_3m_-\right)y\right\}\bu_n(0)
\end{equation}
which can be rewritten as
\begin{equation}
\left(
\begin{array}{c}
u_n^1(y)
\\[6pt]
u_n^2(y)
\end{array}
\right)
\propto
\left(
\begin{array}{c}
\cos\left(\mu_n y\right)
+\frac{m_-}{\mu_n}
\sin\left(\mu_n y\right)
\\[6pt]
\frac{M_n-m_+}{\mu_n}
\sin\left(\mu_n y\right)
\end{array}
\right)
\label{uconstantmm}
\end{equation}
where
\begin{equation}
\mu_n=\sqrt{(M_n-m_+)^2-m_-^2}
\label{mun}
\end{equation}
and we used the boundary condition at $y=0$, eq.\ (\ref{boundcond1}).
The solution (\ref{uconstantmm}) is valid even for imaginary $\mu_n$
(then trigonometrical functions are replaced with appropriate
hyperbolic ones). Now we have to implement also the second
boundary condition, the one at $y=\pm\pi R$. This condition can not be
fulfilled  for imaginary $\mu_n$ (with one exception which we discuss
later). For real $\mu_n$ we can find that the masses
$M_n$ are given by the solution of the following equation
\begin{equation}
(M_n-m_+)\tan(\pi\al)
=
\mu_n
\tan\left[\mu_n\pi R
-
\arctan\left(\frac{m_-}{\mu_n}\right)
\right]
+m_-
\label{massMmm}
\end{equation}
in which $M_n$ appears implicitly on the r.h.s.\ via the parameters
$\mu_n$ defined in eq.\ (\ref{mun}). The same condition for mass
eigenvalues can be obtained from equation (\ref{difft}) with boundary
conditions (\ref{boundcond1t})--(\ref{boundcond2t}).
The reality condition for $\mu_n$ means that the mass eigenstates are
given by (\ref{uconstantmm}) with mass eigenvalues satisfying
(\ref{massMmm}) when
\begin{equation}
\left|M_n-m_+\right| > \left|m_-\right|
\,.
\end{equation}

Now we discuss the additional solution which corresponds to one
exception from the reality condition for $\mu_n$ mentioned above.
It exists only when the Scherk--Schwarz twist is trivial:
$\al=0$. In such a case there is a state with mass $M=m_+$ and shape
given by
\begin{equation}
\left(
\begin{array}{c}
u^1(y)\\[6pt]u^2(y)
\end{array}
\right)
\propto
\left(
\begin{array}{c}
\exp\left\{\int_0^y\d s\,m_-(s)\right\}\\[6pt]0
\end{array}
\right)
.
\end{equation}
Observe that we have used an integral of $m_-(y)$, and not 
just the product of a constant $m_-$ with $y$, in the above
formula. It is not difficult to check that the solution of such  
form is valid for arbitrary $y$--dependent mass term $m_-(y)$.
It is quite important because it describes the massless mode
in the case of vanishing $m_+$. It should be stressed that such a
state appears only when $\al=0$. This means that in the case of a non
trivial Scherk--Schwarz twist there can be no massless mode 
for arbitrary $m_-(y)$ if $m_+=0$.

Let us now go back to the mass eigenvalue condition (\ref{massMmm})
and the mass eigenstates (\ref{uconstantmm}). They allow to extend
our discussion of (non)equivalence of the Scherk--Schwarz mechanism
and direct 5--dimensional mass terms. 
When considering the situation with vanishing $\ZZ_2$--odd mass $m_-$
(see discussion after eqs.\ (\ref{alphaplus}) and (\ref{unchangedKK}))
we have observed that in such a case the Scherk--Schwarz mechanism is
equivalent to adding a constant $\ZZ_2$--even mass $m_+$. Moreover, by
a suitable choice of the Scherk--Schwarz 
twist parameter $\al$ it is possible to reproduce the spectrum (but
not the shapes of the mass eigenstates) obtained in the case of
vanishing $\al$ and given $y$--dependent $m_+$.
The situation changes for $m_-\ne0$ (also if it is constant). From
eq.\ (\ref{massMmm}) it is clear that the masses in the case of the
Scherk--Schwarz mechanism are, for arbitrary $\al$, different from
the masses obtained in a theory with direct mass terms. 
The twist parameter
$\al$ and the $\ZZ_2$--even mass $m_+$ enter the mass formula
(\ref{massMmm}) in quite different ways.
The relation between $m_+$ and mass eigenvalues is very simple.
The constant $m_+$ appears only in the combination $(M_n-m_+)$ so it
just shifts all the mass states by a constant. 
On the other hand, the relation between $\al$ and the mass spectrum
is quite complicated and in general can not be solved explicitly. 
But one feature of that relation can be read from the mass formula
(\ref{massMmm}). For arbitrary $m_+$ and $m_-$ it is possible to
choose $\al$ in such a way that one state with any given mass (within
some range of values) is present
in the spectrum. A state with mass $M_0$ appears in the spectrum
if the Scherk--Schwarz twist parameter is given by
\begin{equation}
\al
=
\frac{1}{\pi}
\arctan\left\{\frac{\mu_0}{M_0-m_+}
\tan\left[\mu_0\pi R-\arctan\left(\frac{m_-}{\mu_0}\right)
\right]+m_-\right\}
,
\end{equation}
with $\mu_0=\sqrt{(M_0-m_+)^2-m_+^2}$ if this mass satisfies the
condition $|M_0-m_+|>|m_-|$. In particular there is a value of $\al$
for which one state is massless if $|m_+|>|m_-|$.

%%%%%%%%%%%%%%%%%%%%%%%%%%%%%%%%%%%%%%%%%%%%%%%%%%%%%%%%%%%%%%%%%%%%%%
%%%%%%%%%%%%%%%%%%%%%%%%%%%%%%%%%%%%%%%%%%%%%%%%%%%%%%%%%%%%%%%%%%%%%%
\subsection{Delta like $\boldsymbol{m_+}$ in the presence of arbitrary
$\boldsymbol{m_-(y)}$}

So far we discussed three special situations: arbitrary $m_+$ with
vanishing $m_-$; constant $m_+$ and $m_-$; arbitrary $m_-$ with 
vanishing $m_+$ (only the massless mode). In general it is not
possible to find explicit solutions for more complicated cases. 
However, there is a very interesting class of models which can be
analysed using the results obtained so far. Let us consider a
situation when the $\ZZ_2$ even and odd mass terms are described by
some arbitrary functions of $y$ subject to the constraint
$m_+(y)m_-(y)=0$ for all $y$. This condition is fulfilled for example 
in all models in which the $\ZZ_2$--even mass terms are located at the
branes. The reason is that arbitrary $m_-(y)$ vanishes at $y=0$ and
$y=\pi R$.

For such models it is possible to check whether a massless state
exists in the spectrum. It is very important because this way one can
determine under what conditions supersymmetry remains unbroken in the
presence of the Scherk--Schwarz twists and the direct mass terms.

First we divide the interval $[0,\pi R]$ into pieces in such a way
that at each piece only one of the mass terms, $m_+(y)$ or $m_-(y)$,
is non vanishing. We find a possible solution corresponding to $M=0$
for each of these sub--intervals. The solution for $y_i<y<y_{i+1}$ is
given by 
\begin{eqnarray}
\bu_0(y)=\exp\left\{i\si_2\int_{y_i}^y\d s\,m_+(s)\right\}\bu_0(y_i)
\,,\quad&&{\rm if\ \ } m_-(y)=0 {\rm\ \ for\ \ } y\in[y_i,y_{i+1}]
\,,\,\,\,\,
\label{step1}
\\[6pt]
\bu_0(y)=\exp\left\{\si_3\int_{y_i}^y\d s\,m_-(s)\right\}\bu_0(y_i)
\,,\,\,\quad&&{\rm if\ \ } m_+(y)=0 {\rm\ \ for\ \ } y\in[y_i,y_{i+1}]
\,.\,\,\,\,
\label{step2}
\end{eqnarray}
We build the full solution from those partial ones starting from the
boundary condition at $y=0$ (\ref{boundcond1}) and demanding that
it is continuous at points $y_i$ where the character of the solution
changes from (\ref{step1}) to (\ref{step2}) or vice versa. At the end
we check whether $\bu_0(\pi R)$ obtained this way, fulfills the
boundary condition (\ref{boundcond2}). Such a procedure allows us to
check what value of the Scherk--Schwarz twist angle is compatible with
a massless fermion for given $y$--dependent mass terms $m_+(y)$ and
$m_-(y)$.

Let us illustrate this by an example of $m_+(y)$ terms located at the
branes: 
\begin{equation}
m_+(y)
=
2m_0\de(y)+2m_\pi\de(y-\pi R)
\,.
\end{equation}
All we have to know about $m_-(y)$ is its integral
\begin{equation}
M_-=\int_0^{\pi R}\d y\,m_-(y)
\,.
\end{equation}
We divide the interval $[0,\pi R]$ into three regions (two of them
infinitesimally small close to the branes) and use the procedure
described above. The result for $\bu_0(\pi R)$ reads
\begin{eqnarray}
\bu_0(\pi R)
\!\!\!&=&\!\!\!
\exp\left\{i\si_2m_\pi\right\}
\exp\left\{\si_3M_-\right\}
\exp\left\{i\si_2m_0\right\}\bu_0(0)
\nn\\[6pt]
\!\!\!&=&\!\!\!
\left(\begin{array}{rr}
\cos(m_\pi)&\sin(m_\pi)\\[6pt]
-\sin(m_\pi)&\cos(m_\pi)
\end{array}\right)
\left(\begin{array}{ll}
e^{M_-}&0\\
0&e^{-M_-}
\end{array}\right)
\left(\begin{array}{rr}
\cos(m_0)&\sin(m_0)\\[6pt]
-\sin(m_0)&\cos(m_0)
\end{array}\right)
\left(\begin{array}{c}
c\\[6pt]0
\end{array}\right)
\nn\\[6pt]
\!\!\!&=&\!\!\!
c\left(\begin{array}{r}
\cos(m_\pi)\cos(m_0)e^{M_-}-\sin(m_\pi)\sin(m_0)e^{-M_-}
\\[6pt]
-\sin(m_\pi)\cos(m_0)e^{M_-}-\cos(m_\pi)\sin(m_0)e^{-M_-}
\end{array}\right)
.
\end{eqnarray}
This solution agrees with the Scherk--Schwarz twist condition
(\ref{boundcond2}) if
\begin{equation}
\frac{
\sin(m_\pi)\cos(m_0)e^{M_-}+\cos(m_\pi)\sin(m_0)e^{-M_-}
}{
\cos(m_\pi)\cos(m_0)e^{M_-}-\sin(m_\pi)\sin(m_0)e^{-M_-}
}
=
-\tan(\pi\al)
\,.
\label{mass0}
\end{equation}
In the case of the gravitino this formula -- relating the mass terms
located at each of the branes, the bulk ($\ZZ_2$--odd) mass term and
the Scherk--Schwarz twist -- must be fulfilled if supersymmetry in the
effective 4--dimensional theory is to be unbroken.

The above formula can be applied also in the case of ordinary
compactification without the Scherk--Schwarz mechanism. Putting $\al$
to zero we obtain the condition
\begin{equation}
\frac{\tan(m_\pi)}{\tan(m_0)}=-e^{-2M_-}
\,.
\label{susy}
\end{equation}
When the $\ZZ_2$--odd mass term integrates to zero, $M_-=0$, we get
the usual condition for the unbroken supersymmetry: $m_\pi=-m_0$ (up
to terms which are multiples of $2\pi$ which we discussed after eq.\
(\ref{unchangedKK})). In the case of vanishing $M_-$ the brane located
mass terms generated for the gravitino must add up to zero. In the
presence of non zero $M_-$ the situation is more complicated and
supersymmetry can be unbroken when the condition (\ref{susy}) is
satisfied.

The condition (\ref{mass0}) for the existence of a massless mode is a
quite complicated relation between the Scherk--Schwarz twist angle
and the three mass parameters: $m_0$, $m_\pi$, $M_-$. Let us now
consider a situation when all the mass parameters are small. Expanding
eq.\ (\ref{mass0}) and keeping only the leading terms we get a much
simpler relation 
\begin{equation}
m_0+m_\pi=-\tan(\pi\al)
\,.
\label{mass0approx}
\end{equation}
Observe that the integral of the $\ZZ_2$--odd mass, $M_-$, dropped out
from the formula. Moreover, in the case of ordinary compactification
($\al=0$) we get a very simple equality $m_\pi=-m_0$ for arbitrary but
small $M_-$. These approximate conditions should be used e.g.\ in
cases when one does not know the full theory but only some terms of an
expansion in some small parameter.

Let us illustrate this with the heterotic M--theory. The field
theoretical limit of this theory is known only up to the first order
of the expansion in $\ka^{2/3}$ (related to the 11--dimensional
coupling constant). We have shown in \cite{Meissner:1999ja} that
effectively there are two kinds of 5--dimensional mass terms for the 
gravitino: the $\ZZ_2$--even one generated by $\left<G_{11abc}\right>$
and the $\ZZ_2$--odd one coming from 
$\left<G_{a\bar ab\bar b}\right>$. Both of them are linear in the
expansion parameter $\ka^{2/3}$ so one should use the approximate
formula (\ref{mass0approx}) and not the full one (\ref{susy}).
Thus the condition for unbroken supersymmetry is just that the brane
located mass terms should sum up to zero as was shown in
\cite{Meissner:1999ja}.

It is possible to gain a qualitative understanding why in more
complicated models the condition for unbroken supersymmetry
(\ref{susy}) differs form the simple global cancellation of the
sources, $m_\pi+m_0=0$, even without the Scherk--Schwarz twist.  
In the presence of the non zero mass term $M_-$ the massless gravitino
is not a constant mode even without any brane terms. In general its
amplitude is different at each brane. Thus, the gravitino 
couples with different strength to sources present at each of the
branes. In general, a simple algebraic cancellation of sources is not
enough to leave supersymmetry unbroken. From eq.\ (\ref{susy}) it is
clear that the details of the shape of the zero mode -- determined by
the details of the function $m_-(y)$ -- are not important. Important
is only the relation between the gravitino wave function at both
branes -- determined by the integral of $m_-(y)$.

%%%%%%%%%%%%%%%%%%%%%%%%%%%%%%%%%%%%%%%%%%%%%%%%%%%%%%%%%%%%%%%%%%%%%%
%%%%%%%%%%%%%%%%%%%%%%%%%%%%%%%%%%%%%%%%%%%%%%%%%%%%%%%%%%%%%%%%%%%%%%
%%%%%%%%%%%%%%%%%%%%%%%%%%%%%%%%%%%%%%%%%%%%%%%%%%%%%%%%%%%%%%%%%%%%%%
\section{Conclusions}

As we have seen, the consistency conditions for a Scherk--Schwarz
mechanism on orbifolds are available in general form  
(subsection \ref{generalcase}). The structure of possible
Scherk--Schwarz twists which can be applied on orbifolds is more
restricted than in the case of manifolds. We found the most general
form of such twists which can be consistently used in 5--dimensional
theories compactified on $S^1/\ZZ_2$. Only rotations in 2--dimensional
subspaces (each consisting of one field with positive and one with
negative parity under $\ZZ_2$) or multiplication by $-1$ are allowed.

The consistency conditions can be used to obtain
differential equations and boundary conditions which determine the
spectrum of the effective theory after compactification.
Solutions, however, are difficult to obtain in closed form due to the
complexity of the problem. We have therefore used a simple $S^1/\ZZ_2$
orbifold to illustrate our results. One of the important ingredients
in the discussion is the general form of the possible mass terms (and
their dependence on the higher dimensional coordinates) that make the
connection to brane induced supersymmetry breaking. Of particular
importance is the appearance of the mass term $m_-$ in eq.\
(\ref{Lkin}). Its existence depends on the presence of VEVs of higher 
dimensional background fields, as was discussed already in
\cite{Meissner:1999ja}. The investigation in
\cite{Bagger:2001qi,Bagger:2001ep} did not consider such a term. 
In addition, their ansatz for the brane located mass terms and
interactions differs from the one adopted here.

Explicit solutions have been presented for two classes of models:
those with constant mass terms $m_+$ and $m_-$ and those with
arbitrary $y$--dependent $m_+(y)$ but with vanishing $m_-$. 
Closed solutions for general mass terms $m_+(y)$ and $m_-(y)$ are
difficult to obtain. Fortunately some of the most interesting cases,
as e.g those with delta--like mass terms located at the branes can be
analysed with sufficient accuracy. It is possible to find conditions
determining whether massless fermions are present in the
spectrum. This allows to check if supersymmetry is broken in a theory
with given form of the mass terms $m_+(y)$ and $m_-(y)$ in the
presence of non trivial Scherk--Schwarz twists.

%%%%%%%%%%%%%%%%%%%%%%%%%%%%%%%%%%%%%%%%%%%%%%%%%%%%%%%%%%%%%%%%%%%%%%
%%%%%%%%%%%%%%%%%%%%%%%%%%%%%%%%%%%%%%%%%%%%%%%%%%%%%%%%%%%%%%%%%%%%%%
%%%%%%%%%%%%%%%%%%%%%%%%%%%%%%%%%%%%%%%%%%%%%%%%%%%%%%%%%%%%%%%%%%%%%%
\section*{Acknowledgments}

We would like to thank S. Groot Nibbelink for useful discussions. 
\\ 
Work supported in part by the European Community's Human Potential
Programme under contracts HPRN--CT--2000--00131 Quantum Spacetime,
HPRN--CT--2000--00148 Physics Across the Present Energy Frontier
and HPRN--CT--2000--00152 Supersymmetry and the Early Universe.
KM and MO were partially supported by the Polish KBN grant 
2 P03B 052 16.

\vspace{2cm}
%%%%%%%%%%%%%%%%%%%%%%%%%%%%%%%%%%%%%%%%%%%%%%%%%%%%%%%%%%%%%%%%%%%%%%
%%%%%%%%%%%%%%%%%%%%%%%%%%%%%%%%%%%%%%%%%%%%%%%%%%%%%%%%%%%%%%%%%%%%%%
%%%%%%%%%%%%%%%%%%%%%%%%%%%%%%%%%%%%%%%%%%%%%%%%%%%%%%%%%%%%%%%%%%%%%%


\begin{thebibliography}{99}

%\cite{Scherk:1978ta}
\bibitem{Scherk:1978ta}
J.~Scherk and J.H.~Schwarz,
%``Spontaneous Breaking Of Supersymmetry Through Dimensional Reduction,''
Phys.\ Lett.\ B {\bf 82} (1979) 60.
%%CITATION = PHLTA,B82,60;%%

%\cite{Scherk:1979zr}
\bibitem{Scherk:1979zr}
J.~Scherk and J.H.~Schwarz,
%``How To Get Masses From Extra Dimensions,''
Nucl.\ Phys.\ B {\bf 153} (1979) 61.
%%CITATION = NUPHA,B153,61;%%

%\cite{Rohm:aq}
\bibitem{Rohm:aq}
R.~Rohm,
%``Spontaneous Supersymmetry Breaking In Supersymmetric String Theories,''
Nucl.\ Phys.\ B {\bf 237} (1984) 553.
%%CITATION = NUPHA,B237,553;%%

%\cite{Dixon:jw}
\bibitem{Dixon:jw}
L.J.~Dixon, J.A.~Harvey, C.~Vafa and E.~Witten,
%``Strings On Orbifolds,''
Nucl.\ Phys.\ B {\bf 261} (1985) 678.
%%CITATION = NUPHA,B261,678;%%

%\cite{Dixon:1986jc}
\bibitem{Dixon:1986jc}
L.J.~Dixon, J.A.~Harvey, C.~Vafa and E.~Witten,
%``Strings On Orbifolds. 2,''
Nucl.\ Phys.\ B {\bf 274} (1986) 285.
%%CITATION = NUPHA,B274,285;%%

%\cite{Candelas:en}
\bibitem{Candelas:en}
P.~Candelas, G.T.~Horowitz, A.~Strominger and E.~Witten,
%``Vacuum Configurations For Superstrings,''
Nucl.\ Phys.\ B {\bf 258} (1985) 46.
%%CITATION = NUPHA,B258,46;%%

%\cite{Horava:1996vs}
\bibitem{Horava:1996vs}
P.~Ho\v rava,
%``Gluino condensation in strongly coupled heterotic string theory,''
Phys.\ Rev.\ D {\bf 54} (1996) 7561
[arXiv:hep-th/9608019].
%%CITATION = HEP-TH 9608019;%%

%\cite{Nilles:1997cm}
\bibitem{Nilles:1997cm}
H.P.~Nilles, M.~Olechowski and M.~Yamaguchi,
%``Supersymmetry breaking and soft terms in M-theory,''
Phys.\ Lett.\ B {\bf 415} (1997) 24
[arXiv:hep-th/9707143].
%%CITATION = HEP-TH 9707143;%% 

%\cite{Antoniadis:1997ic}
\bibitem{Antoniadis:1997ic}
I.~Antoniadis and M.~Quiros,
%``Supersymmetry breaking in M-theory and gaugino condensation,''
Nucl.\ Phys.\ B {\bf 505} (1997) 109
[arXiv:hep-th/9705037].
%%CITATION = HEP-TH 9705037;%%

%\cite{Antoniadis:1997xk}
\bibitem{Antoniadis:1997xk}
I.~Antoniadis and M.~Quiros,
%``On the M-theory description of gaugino condensation,''
Phys.\ Lett.\ B {\bf 416} (1998) 327
[arXiv:hep-th/9707208].
%%CITATION = HEP-TH 9707208;%%

%\cite{Nilles:1998sx}
\bibitem{Nilles:1998sx}
H.P.~Nilles, M.~Olechowski and M.~Yamaguchi,
%``Supersymmetry breakdown at a hidden wall,''
Nucl.\ Phys.\ B {\bf 530} (1998) 43
[arXiv:hep-th/9801030].
%%CITATION = HEP-TH 9801030;%%

%\cite{Meissner:1999ja}
\bibitem{Meissner:1999ja}
K.A.~Meissner, H.P.~Nilles and M.~Olechowski,
%``Supersymmetry breakdown at distant branes: The super-Higgs mechanism,''
Nucl.\ Phys.\ B {\bf 561} (1999) 30
[arXiv:hep-th/9905139].
%%CITATION = HEP-TH 9905139;%%

%\cite{Horava:1996ma}
\bibitem{Horava:1996ma}
P.~Ho\v rava and E.~Witten,
%``Eleven-Dimensional Supergravity on a Manifold with Boundary,''
Nucl.\ Phys.\ B {\bf 475} (1996) 94
[arXiv:hep-th/9603142].
%%CITATION = HEP-TH 9603142;%%

%\cite{Bagger:2001qi}
\bibitem{Bagger:2001qi}
J.A.~Bagger, F.~Feruglio and F.~Zwirner,
%``Generalized symmetry breaking on orbifolds,''
Phys.\ Rev.\ Lett.\  {\bf 88} (2002) 101601
[arXiv:hep-th/0107128].
%%CITATION = HEP-TH 0107128;%%

%\cite{Bagger:2001ep}
\bibitem{Bagger:2001ep}
J.~Bagger, F.~Feruglio and F.~Zwirner,
%``Brane induced supersymmetry breaking,''
JHEP {\bf 0202} (2002) 010
[arXiv:hep-th/0108010].
%%CITATION = HEP-TH 0108010;%%

%\cite{Hosotani:1983xw}
\bibitem{Hosotani:1983xw}
Y.~Hosotani,
%``Dynamical Mass Generation By Compact Extra Dimensions,''
Phys.\ Lett.\ B {\bf 126} (1983) 309.
%%CITATION = PHLTA,B126,309;%%

%\cite{Hosotani:1988bm}
\bibitem{Hosotani:1988bm}
Y.~Hosotani,
%``Dynamics Of Nonintegrable Phases And Gauge Symmetry Breaking,''
Annals Phys.\  {\bf 190} (1989) 233.
%%CITATION = APNYA,190,233;%%

\end{thebibliography}
\end{document}